\documentstyle[12pt]{article}

\newcommand{\be}{\begin{equation}}
\newcommand{\ee}{\end{equation}}
 
\newcommand{\bea}{\begin{eqnarray}}
\newcommand{\eea}{\end{eqnarray}}

\newcommand{\NP}[1]{Nucl. Phys.\ {\bf #1}\ }
\newcommand{\PL}[1]{Phys. Lett.\ {\bf #1}\ }

\newcommand{\PR}[1]{Phys. Rev.\ {\bf #1}\ }
\newcommand{\PRL}[1]{Phys. Rev. Lett.\ {\bf #1}\ }

\newcommand{\MPL}[1]{Mod. Phys. Lett.\ {\bf #1}\ }

\def\lsim{\raise0.3ex\hbox{$<$\kern-0.75em\raise-1.1ex\hbox{$\sim$}}}
\def\gsim{\raise0.3ex\hbox{$>$\kern-0.75em\raise-1.1ex\hbox{$\sim$}}}

\newcommand{\meg}{\mu\rightarrow e\gamma}

\renewcommand{\ln}{{\rm ln}}

\begin{document}

\begin{titlepage}
\begin{flushright}
\parbox[t]{4.8truecm}{
\begin{center}
{\large  HU-SEFT R 1996-09\\}
%{(hep-ph/9603412)}
\end{center}}
\end{flushright}
\vskip 2cm
\begin{center}
\vskip .2in
 
{\Large \bf
Constraints on $R$-parity violating interactions from
$\mu\rightarrow e\gamma$}

\vskip .4in
 
Masud Chaichian$^{a,b}$\footnote{
Email: chaichian@phcu.helsinki.fi} and 
Katri Huitu$^{a}$\footnote{
Email: huitu@phcu.helsinki.fi}\\[.15in]
 
{\em $^{a}$Research Institute for High Energy Physics}\\[.15in]
 
{\em $^{b}$High Energy Physics Division,
     Department of Physics}\\[.15in]
 
{\em      
P.O. Box 9 (Siltavuorenpenger 20 C)\\
FIN-00014 University of Helsinki, Finland}
 
\end{center}
 
\vskip 1.5in
 
\begin{abstract}
 
The lepton number violating process $\mu\rightarrow
e\gamma$ is used to study the lepton number violating couplings
in MSSM extended by terms violating $R$-parity explicitly.
Bounds are obtained for the products of $\lambda $- or $\lambda '$-type
couplings.
It is found that many of these limits are more stringent
than the ones obtained previously.

\end{abstract}

\vskip 1.truecm

\begin{flushleft}
PACS numbers: 12.60.Jv, 11.30.Fs, 14.80.Ly, 11.30.Qc
\end{flushleft}

\vskip 1.truecm

\end{titlepage}

\newpage
\renewcommand{\thepage}{\arabic{page}}
\setcounter{page}{1}
\setcounter{footnote}{1}
 
The decay of muon to electron and photon is a useful process to
test theories with lepton number violating interactions,
since it is experimentally very strictly constrained.
In the context of supersymmetry, this process has been studied earlier 
in the Minimal Supersymmetric Standard Model (MSSM) assuming either
universality of the soft scalar masses at the GUT scale \cite{KLV} 
or letting the non-diagonal scalar masses or trilinear soft
couplings to be arbitrary \cite{CEKLP}.
In the first case it was found that the non-universality coming from
the RGE evolution from GUT to weak scale is too small to be observed,
while in the second case the larger the non-diagonal terms at the
GUT scale, the heavier the spectrum of the supersymmetric partners should 
be to satistfy the experimental bounds.
Here we will assume that the contribution from mixing of the
slepton generations is negligible and look for other possible sources 
of lepton number violation.

In the MSSM, the conservation of lepton and baryon number in the 
Standard Model has been put in by conservation of 
the so-called $R$-parity, $R=(-1)^{3(B-L)+2S}$.
The supersymmetry and gauge invariance allow also $R$-violating
interactions, namely

\be
W_{R\!\!\!\!/} = \lambda_{ijk} \widehat L_i\widehat L_j\widehat E^c_k +
\lambda '_{ijk} \widehat L_i\widehat Q_j\widehat D^c_k +
\lambda ''_{ijk} \widehat U^c_i\widehat D^c_j\widehat D^c_k
-\mu_i L_i H_2.
\label{WRviol}
\ee

\noindent
Here $\lambda_{ijk} = -\lambda_{jik} $ and 
$\lambda ''_{ijk} = -\lambda ''_{ikj} $.
The $\lambda,\,\lambda '$ and $\mu  $ terms
violate the lepton number, whereas the $\lambda '' $ terms
violate the baryon number by one unit.
We assume that the $\mu $ terms can be rotated away by a 
redefinition of fields \cite{HS,DP}.
It is worth noting that models exist, in which the violation of
$R$-parity is a necessity, e.g. in the supersymmetric left-right model
the $R$-parity is automatically conserved at the level of
the Lagrangian, but in the genuine minimum of the potential for this 
model one or more of the
sneutrinos get a VEV and hence violate the lepton number
and $R$-parity \cite{KM}.
Simultaneous presence of $L$ violating ($\lambda_{ijk}, 
\lambda '_{ijk} $) and $B$ violating ($\lambda ''_{ijk} $) couplings
would generally lead to too fast proton decay.
It was recently argued \cite{SV} that proton decay for squark masses below 
1 TeV always constrains the product $|\lambda '\cdot\lambda ''| < 10^{-9}$.
Huge difference between the strengths of lepton and baryon number
violating couplings are also supported by
examples of GUT models, in which quarks and leptons are treated
differently \cite{HS,Hempfling}.
Furthermore, it is well known that in string unification the conservation 
of $R$-parity is not evident \cite{BHRIR}. 
In this paper it is assumed that only the lepton number violating
couplings are non-vanishing.

The strength of the couplings in Eq. (\ref{WRviol}) has been studied 
from several different sources, e.g. charged current universality, 
$e-\mu -\tau$ universality, forward-backward asymmetry,
$\nu_\mu e$ scattering, atomic parity violation \cite{BGH},
neutrinoless double beta decay \cite{HKK}, $\nu $ masses \cite{GRT}, 
heavy nuclei decay \cite{BMDH}, $Z$-boson partial width \cite{Bhatta1}, 
and $K^+,\, t$-quark decays \cite{AG}.
We have updated a table of limits for $\lambda $- and $\lambda '$ -couplings,
Table \ref{prevlim}.
The bounds are typically between $10^{-3} - 10^{-1}$ as seen from Table 
\ref{prevlim}.
Limits on neutrino masses give stringent bounds for $\lambda_{133}$
and $\lambda_{133} '$ couplings due to tau lepton or bottom quark in
the loop inducing Majorana mass term \cite{GRT}.

Severe limitations come also from baryogenesis considerations if
it is required that the primordial baryon asymmetry is preserved,
since $L$ violating interactions in equilibrium together with $B+L$ 
violating sphalerons would wash out any pre-existing baryon asymmetry.
The sphalerons preserve $\frac 13 B-L_i$ and it has been shown in
\cite{DR} that any bound from cosmology can be avoided
by demanding that one of the lepton numbers is 
conserved.
Since we wish to study the decay $\meg $,
we could have conservation of $\tau $ lepton number.
On the other hand, if electroweak baryogenesis is assumed (see,
e.g. \cite{FY}), the restrictions on lepton number are removed.
We shall give our results both assuming $\tau $ number conservation
and relaxing this assumption.

\begin{table}
\begin{tabular}{|c|c|c|c|c|c|}
\hline 
$\lambda_{ijk}<$ & $m_{\tilde f}{=100{\rm GeV}} $ &
$\lambda '_{ijk}<$ & $m_{\tilde f}{=100{\rm GeV}} $ &
$\lambda '_{ijk}<$ & $m_{\tilde f}{=100{\rm GeV}} $ \\
\hline
12k & 0.04 \cite{BGH} & 111 & 0.0004 \cite{HKK} & 22k & 0.012 \cite{AG} \\
131 & 0.10 \cite{BGH} & 112 & 0.03 \cite{BGH} &  231 & 0.22 \cite{BGH}\\
132 & 0.10 \cite{BGH} & 113 & 0.03 \cite{BGH} & 232& 0.44 \cite{Bhatta1} \\
133 & 0.001 \cite{GRT} & 12k & 0.012 \cite{AG} & 233 & 0.44 \cite{Bhatta1} \\
23k & 0.09 \cite{BGH} & 131 & 0.26 \cite{BGH} & 31k & 0.012 \cite{AG}\\ 
 &  & 132 & 0.51 \cite{Bhatta1} & 32k & 0.012 \cite{AG} \\
 & & 133 & 0.001 \cite{GRT} & 33k & 0.26 \cite{Bhatta1}\\
 & & 21k & 0.012 \cite{AG} &  &  \\
\hline
\end{tabular}
\caption{\label{prevlim} Previously found limits on single $\lambda_{ijk} $
and $\lambda '_{ijk}$.}
\end{table}

Addition of Eq. (\ref{WRviol}) to MSSM superpotential leads to more 
interactions in the model, but the MSSM particle content remains.
The relevant part of the Lagrangian
is found by the standard techniques \cite{HK}%,GRT}

\bea
L_{L\!\!\! /\, ,\lambda, \lambda '}&=&
\lambda_{ijk}(\tilde\nu_{iL}\bar e_{kR} e_{jL}
+\tilde e_{jL}\bar e_{kL} \nu_{iL}
+\tilde e_{kR}^*\overline{\nu_{iL}^c} e_{jL}  
-(i\leftrightarrow j))  \nonumber \\
&&+\lambda '_{ijk}(-\tilde u_{jL}\bar d_{kR} e_{iL}
-\tilde d_{kR}^*\overline{e_{iL}^c} u_{jL}) +h.c.,
\eea

\begin{figure}[t]
\leavevmode
\begin{center}
\mbox{\epsfxsize=13.2truecm\epsfysize=13.6truecm\epsffile{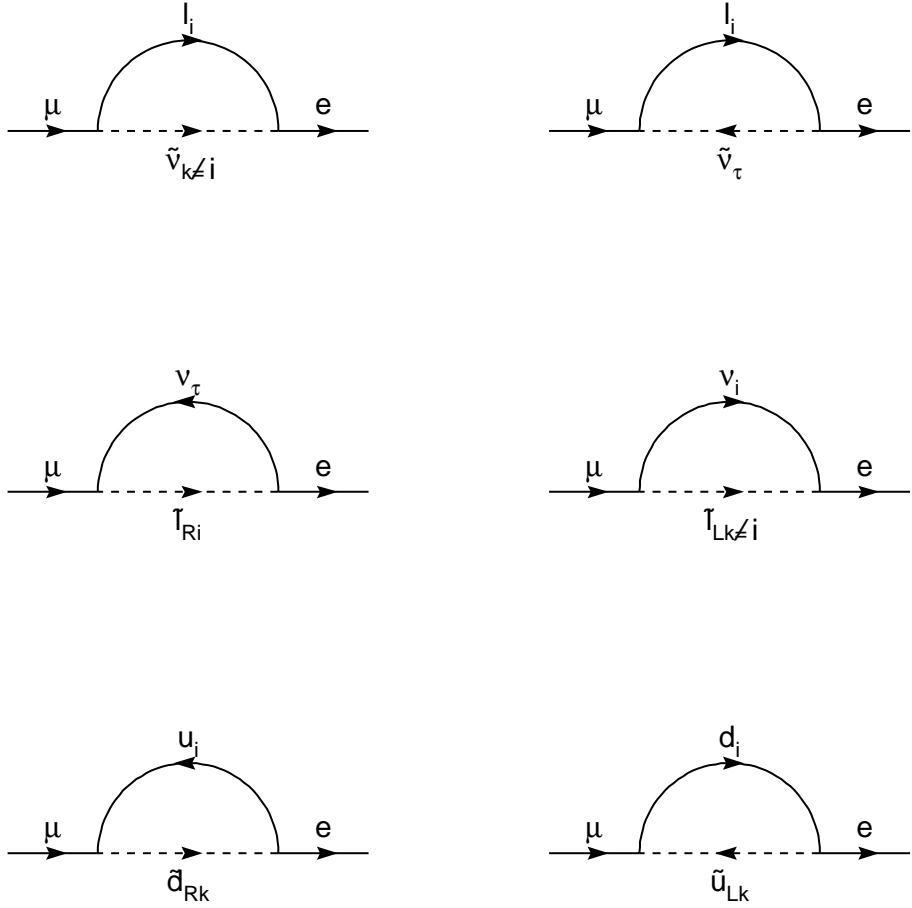}}
\end{center}
\caption{\label{graphs} Contributing diagrams for the muon decay to
electron and photon.
Unless otherwise indicated $i,k=1\ldots 3$.}
\end{figure}

\noindent
{}from which
the contributions to the 
radiative muon decay are found.
These are shown in Fig. (\ref{graphs}).
The photon line is not shown, but it should be attached in all
possible ways to the graphs.
If $\tau $ number conservation is assumed, only those graphs which 
have in the loop none or two particles carrying $\tau $ number should
be included.

\input{epsf.sty}

The gauge invariant amplitude for $\meg $ is usually parametrized as

\be
T(\meg ) =\epsilon^\lambda \bar u_e (p') (A + B\gamma _5) 
i\sigma_{\lambda\nu} q ^\nu u_\mu (p),
\label{amplitude}
\ee

\noindent
where $p,\,p'$ and $q$ are the momenta of muon, electron and photon,
respectively.
$\epsilon^\lambda$ is the photon polarization vector and 
$\sigma_{\lambda\nu}=\frac{i}{2}[\gamma_\lambda,\gamma_\nu]$.
It is easily seen from Eq. (\ref{amplitude}), that the amplitude is 
nonvanishing only when the muon and the electron are of opposite helicities.

The width of the decay can be evaluated using the amplitude, 
Eq. (\ref{amplitude}),
and finally the branching ratio from $\mu $ lifetime,
$\tau_\mu = 192 \pi^2 /(G_F^2 m_\mu^5) $, with the result

\be
BR=\frac {24 \pi }{G_F^2 m_\mu^2} (|A|^2 + |B|^2).
\label{BR}
\ee

\noindent
Similarly one could study the decay of tau lepton to muon and photon
or electron and photon.

The experimental limits for the lepton decays are given as
\cite{PDG}

\bea
BR(\meg ) &<& 4.9\cdot 10^{-11} ,\nonumber\\
BR(\tau \rightarrow \mu\gamma ) &<& 4.2\cdot 10^{-6},\nonumber\\
BR(\tau \rightarrow e\gamma ) &<& 1.2\cdot 10^{-4}.
\eea

Using Gordon decomposition for practical calculations, the relevant 
part of all the amplitudes corresponding to the graphs in
Fig. (\ref{graphs}) are found.
$A$ in the amplitude, Eq. (\ref{amplitude}), can be written in terms of
the following functions:

\bea
A_1 &=& \frac{\lambda_1 \lambda_2 Q m_\mu}{16 \pi^2 m_{\tilde f}^2}
\frac {1}{6 (\kappa -1)^3} \left(-\kappa^2 +5\kappa +2
-\frac{6\kappa }{\kappa -1}\, \ln\kappa \right), \nonumber\\
A_2 &=& \frac{\lambda_1 \lambda_2 Q m_\mu}{16 \pi^2 m_{\tilde f}^2}
\frac {1}{6 (\kappa -1)^3} \left(2 \kappa^2 +5\kappa -1
-\frac{6\kappa^2 }{\kappa -1}\, \ln\kappa \right) .
\eea

\noindent
There are two lepton number violating vertices in every
contribution, characterized by $\lambda_{1,2}=\lambda_{i,j,k} $ 
or $\lambda_{1,2}=\lambda '_{i,j,k}$
couplings.
The functions $A_1$ and $A_2$ depend also on the
charge of the particle attached to the photon $Q$,
the sfermion mass occurring in the loop $m_{\tilde f}$, and the ratio of
the masses of fermion and sfermion in the loop,
$\kappa=m_f^2/\tilde m_f^2$.
Proportionality to the mass of the muon, $m_\mu $, reflects
the helicity flip on the external muon line.
The lack of a term proportional to the electron mass 
on the other hand indicates
that we have approximated the external electron to be massless.
The $A_1$ function corresponds to the situation where the photon line is
attached to the fermion line and $A_2$ to the situation where
the photon line is attached to the scalar.
In all cases $|A|=|B|$ in Eq. (\ref{amplitude}).

We have analyzed three cases, namely
i) the $\lambda $ couplings dominate and are the same, 
$\lambda_{ijk}=\lambda $,
ii) the $\lambda '$ couplings dominate and are the same, 
$\lambda '_{ijk}=\lambda '$,
iii) one pair of the $\lambda $ or $\lambda '$ couplings dominates over
the others.

In cases i) and ii), the spectrum is calculated by assuming universal
scalar mass and universal gaugino mass at the GUT scale $10^{16}$ GeV
and evaluating the masses
down to the electroweak scale \cite{martin}.
Two sets of universal mass parameters are used, one leading to
a light SUSY spectrum (slepton and squark masses between 100--400 GeV) 
and another one leading
to heavy SUSY spectrum with masses 1--1.4 TeV. 
\begin{figure}[t]
\leavevmode
\begin{center}
\mbox{\epsfxsize=12.cm\epsfysize=12.cm\epsffile{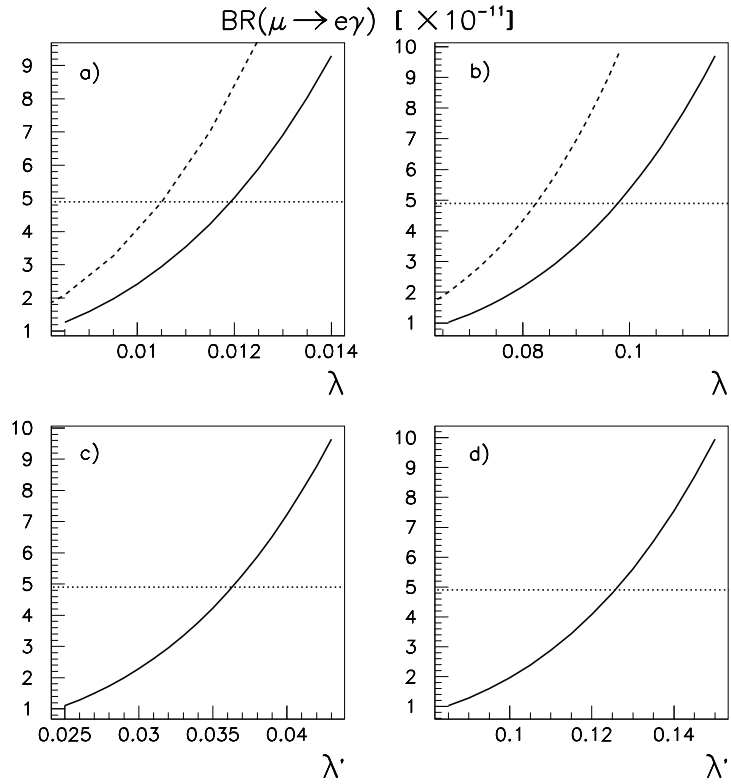}}
\end{center}
\caption{\label{limits}
The branching ratio for $\meg $, when 
a) and b): $\lambda_{ijk} = \lambda $ and $\lambda_{ijk} '=0$ 
for all $i,j,k$
and c) and d): $\lambda_{ijk} '= \lambda '$ and $\lambda_{ijk} =0$ 
for all $i,j,k$. 
In a) and c) the SUSY spectrum is light and in b) and d) the SUSY spectrum
is heavy. 
The dashed lines in a) and b) correspond to the situation in which
$\tau $ number conservation has been relaxed.
The dotted line is the experimental limit, $BR(\meg )<4.9\cdot 10^{-11}$.}
\end{figure}
In Figs. (\ref{limits}) a) and c), the solid lines give the upper limits
for the couplings when $\tau $ number is conserved.
If $\tau $ number conservation is relaxed (dashed lines), more graphs
contribute and limit for the couplings become stricter.
{}From Figs. (\ref{limits}) a) and c) it is seen that when the spectrum is 
light, the maximum coupling is $\cal{O}$$(10^{-2})$.
In the case of the heavy spectrum, Figs. (\ref{limits}) b) and d), the maximum 
coupling is approximately one third of the electromagnetic coupling.

In case iii), a large hierarchy between various pairs of $\lambda $
or $\lambda '$ couplings has been assumed.
Case iii) is in a sense a natural choice, since the $\lambda $ and
$\lambda '$ couplings are similar to the Yukawa couplings, which
are known to vary over at least six orders of magnitude.
This is also the most conservative limit for the couplings.
This case has been studied for two different scalar masses in the loop,
namely $m_{\tilde f} = 100$ GeV and $m_{\tilde f} = 1$ TeV.
For the lighter scalar mass, one sees from Table \ref{opt3}
that the last three limits are much less strict than the others.
This is due to the top quark ($m_{top}\sim 175$ GeV) in the loop,
since then the value of $\kappa$ is larger than one.
Also the effect of the bottom quark is seen in some of the bounds,
especially in the last one, when both top- and bottom-quarks contribute
to the result.
When the scalar mass in the loop is 1 TeV, one does not anymore see
the effect of the bottom quark and also the effect due to the
top quark is much less prominent.

Comparing these products with the earlier limits, Table \ref{prevlim},
it is seen that only if the previous limit comes from the constraint
on the neutrino mass, double $\beta $ decay or $K^+$ meson decays, 
it is more stringent than 
the present one.

\begin{table}
\begin{tabular}{|c|c|c||c|}
\hline
& $m_{\tilde f}= 100$ GeV & $m_{\tilde f}= 100$ GeV &$m_{\tilde f}= 1$ TeV \\ 
&& previous results & \\
& $\times 10^{-4}$ & $\times 10^{-4}$ &  $\times 10^{-2}$ \\
\hline
$|\lambda_{121}\, \lambda_{122}|\,<\,$ & $1.0 $ & 16 & $1.0 $ \\
$|\lambda_{131}\, \lambda_{132}|\,<\,$ & $1.0 $ & 100 & $1.0 $ \\
$|\lambda_{231}\, \lambda_{232}|\,<\,$ & $1.0 $ & 81 & $1.0 $ \\
$|\lambda_{231}\, \lambda_{131}|\,<\,$ & $2.0 $ & 90 & $2.0 $ \\
$|\lambda_{232}\, \lambda_{132}|\,<\,$ & $2.0 $ & 90 & $2.0 $ \\
$|\lambda_{233}\, \lambda_{133}|\,<\,$ & $2.0 $ & 0.9 & $2.0 $ \\
$|\lambda '_{211}\, \lambda '_{111}|\,<\,$ & $8.0 $ & .048 & $8.0 $ \\
$|\lambda '_{212}\, \lambda '_{112}|\,<\,$ & $8.0 $ & 1.4 & $8.0 $ \\
$|\lambda '_{213}\, \lambda '_{113}|\,<\,$ & $8.1 $ & 1.4 & $8.0 $ \\
$|\lambda '_{221}\, \lambda '_{121}|\,<\,$ & $8.1 $ & 1.4 & $8.0 $ \\
$|\lambda '_{222}\, \lambda '_{122}|\,<\,$ & $8.1 $ & 1.4 & $8.0 $ \\
$|\lambda '_{223}\, \lambda '_{123}|\,<\,$ & $8.2 $ & 1.4 & $8.0 $ \\
$|\lambda '_{231}\, \lambda '_{131}|\,<\,$ & $140 $ & 570 & $10.4 $ \\
$|\lambda '_{232}\, \lambda '_{132}|\,<\,$ & $140 $ & 2200 & $10.4 $ \\
$|\lambda '_{233}\, \lambda '_{133}|\,<\,$ & $180 $ & 4.4 & $10.4 $ \\
\hline
\end{tabular}
\caption{\label{opt3}
Upper limits on products of $\lambda_{ijk} $ and 
$\lambda '_{ijk}$ couplings for two different scalar masses, in the first
column $m_{\tilde f}=$100 GeV and in the third $m_{\tilde f}=$1 TeV.
The second column contains earlier results from Table 1.}% \ref{prevlim}.}
\end{table}

If the $\tau $ number were not broken, the limits in
Table \ref{opt3} on $\lambda $ type couplings should be included only 
when none or two of the $i,j,k$ are 3's.
One could also study the limits from $\tau\rightarrow
\mu\gamma$ or $\tau\rightarrow e\gamma$, but as it appears 
these limits would not be tight enough to strengthen the
bounds found previously.

To summarize, we have shown that the experimental upper limit on the
muon radiative decay can be used to obtain stringent bounds on the
magnitude of the $R$-parity violating interactions.
Although we have considered the case of explicit $R$-parity 
breaking, the case of spontaneous breaking of this symmetry can be 
treated by a similar analysis.
The interplay between the dominance of cross sections for reactions with
exact or broken $R$-parity implemented by such bounds, is of great
interest in the search for supersymmetric particles.

\end{document}